# A Mathematical Treatment to Determine Transient Growth Kinetics from a Given Size Distribution

Yanhao Dong (dongyh@mit.edu) and Jian Han

*Department of Materials Science and Engineering, University of Pennsylvania,*

*Philadelphia, PA 19104, USA*

Particle coarsening and grain growth take place to minimize the total interfacial energy. The classical mean-field treatments by Lifshitz, Slyozov, [1] Wagner [2] and Hillert [3] predicted cubic growth law under bulk-diffusion controlled precipitation coarsening and parabolic growth law under interface controlled grain growth, as well as their steady-state size distribution. When the size distribution is the steady-state one, the average grain size $\bar{G}$ satisfies the following dependence: $\bar{G}^3 \sim t$ under bulk-diffusion control and $\bar{G}^2 \sim t$ under interface control, with correct slopes definitely given by the theory. However, when the size distribution does not satisfy the steady-state one, the growth kinetics would deviate from theoretical prediction. As is shown by numerical simulations in Ref. 4 and 5, the deviation is less obvious in the growth law, but more evident in the slope of the growth curve. Specifically, starting from a non-steady-state size distribution, the parabolic/cubic growth law would be recovered quickly, yet the convergence of the slope in the $\bar{G}^3 \sim t$ or $\bar{G}^2 \sim t$ curve is extremely slow. Therefore, it would be interesting to know how to calculate the instantaneous slope of the growth kinetics from an arbitrarily given size distribution. As will be shown below, assuming a known growth law, the analytical solution can be easily obtained and the exact solution can be obtained by numerical integration.

Consider an interface controlled growth [3] and the coarsening equation for a grain with a size $G$ described by

$$\frac{dG}{dt} = 2M_b\gamma\left(\frac{1}{G_{cr}} - \frac{1}{G}\right) \quad (1)$$

where $t$ is the time, $M_b$ is the grain boundary mobility, $\gamma$ is the grain boundary energy and $G_{cr}$ is a critical grain size that neither shrink nor grow at time $t$. By mass conservation, we have

$$\sum G^2 \frac{dG}{dt} = 0 \quad (2)$$

With Eq. (1), it yields

$$G_{cr} = \frac{\sum G^2}{\sum G} \quad (3)$$

The steady-state solution by Hillert gives

$$\frac{d\bar{G}^2}{dt} = \left(\frac{32}{81}\right)(2M_b\gamma) \quad (4)$$

with a normalized size distribution

$$P_{eq}(x) = P_{eq}\left(\frac{G}{\bar{G}}\right) = \frac{3\left(\frac{8}{9}x\right)\cdot(2e)^3}{\left(2-\frac{8}{9}x\right)^5}\exp\left(\frac{-6}{2-\frac{8}{9}x}\right) \quad (5)$$

Now suppose we start with a known normalized size distribution $P\left(\frac{G}{\bar{G}}, t=0\right)$ and seek to solve the transient growth rate $\left.\frac{d\bar{G}^2}{dt}\right|_{t=0}$. We have

$$\bar{G} = \frac{\int GP\left(\frac{G}{\bar{G}},t\right)dG}{\int P\left(\frac{G}{\bar{G}},t\right)dG} = \int GP\left(\frac{G}{\bar{G}},t\right)dG \quad (6)$$

$$\frac{d\bar{G}^2}{dt} = 2\bar{G}\frac{d\bar{G}}{dt}$$

$$= 2\bar{G}\frac{\int G\frac{dP(G/\bar{G},t)}{dt}dG\int P(G/\bar{G},t)dG - \int GP(G/\bar{G},t)dG\int\frac{dP(G/\bar{G},t)}{dt}dG}{\left(\int P\left(\frac{G}{\bar{G}},t\right)dG\right)^2}$$

$$= 2\bar{G}\left[\int G\frac{dP\left(\frac{G}{\bar{G}},t\right)}{dt}dG - \int GP\left(\frac{G}{\bar{G}},t\right)dG\int\frac{dP\left(\frac{G}{\bar{G}},t\right)}{dt}dG\right]$$

(7)

With the continuity equation in the grain size space

$$\frac{\partial P}{\partial t} + \frac{\partial}{\partial G}\left(P\frac{dG}{dt}\right) = 0 \qquad (8)$$

Eq. (7) reduces to

$$\frac{d\bar{G}^2}{dt} = 2\bar{G}\left[\int G\frac{dP\left(\frac{G}{\bar{G}},t\right)}{dt}dG + \int GP\left(\frac{G}{\bar{G}},t\right)dG\int\frac{\partial}{\partial G}\left(P\left(\frac{G}{\bar{G}},t\right)\frac{dG}{dt}\right)dG\right]$$

$$= 2\bar{G}\left[-\int G\frac{\partial}{\partial G}\left(P\left(\frac{G}{\bar{G}},t\right)\frac{dG}{dt}\right)dG + \int GP\left(\frac{G}{\bar{G}},t\right)dG\left(P\left(\frac{G}{\bar{G}},t\right)\frac{dG}{dt}\right)\Big|_{G_{\min}}^{G_{\max}}\right]$$

(9)

Integrating the first part in the bracket by parts, we obtain

$$\frac{d\bar{G}^2}{dt} = 2\bar{G}\left[-\left(GP\left(\frac{G}{\bar{G}},t\right)\frac{dG}{dt}\right)\Big|_{G_{\min}}^{G_{\max}} + \int P\left(\frac{G}{\bar{G}},t\right)\frac{dG}{dt}dG + \int GP\left(\frac{G}{\bar{G}},t\right)dG\left(P\left(\frac{G}{\bar{G}},t\right)\frac{dG}{dt}\right)\Big|_{G_{\min}}^{G_{\max}}\right]$$

(10)

where

$$\frac{dG}{dt} = 2M_b\gamma\left(\frac{\int GP\left(\frac{G}{\bar{G}},t\right)dG}{\int G^2P\left(\frac{G}{\bar{G}},t\right)dG} - \frac{1}{G}\right) \qquad (11)$$

Therefore, $\dfrac{d\bar{G}^2}{dt}$ is now written as a function that only depends on $P\left(\dfrac{G}{\bar{G}},t\right)$ and the solution of $\left.\dfrac{d\bar{G}^2}{dt}\right|_{t=0}$ is complete. To check the consistency, we let $P\left(\dfrac{G}{\bar{G}},t\right) = P_{eq}\left(\dfrac{G}{\bar{G}}\right)$ and $\left.\dfrac{d\bar{G}^2}{dt}\right|_{t=0}$ returns a value of $0.395(2M_b\gamma)$, which agrees well the predicted value of $\left(\dfrac{32}{81}\right)(2M_b\gamma)$ by Hillert. Therefore, the same coarsening kinetics of Eq. (4) has been confirmed.

The above approach equally applies to growth problem governed by other coarsening equations, such as the diffusion controlled growth which was solved by Lifshitz, Slyozov and Wagner (LSW) [1, 2]. The coarsening equation can be written as

$$\frac{dG}{dt} = \frac{2D\Omega\gamma}{G}\left(\frac{1}{G_{cr}} - \frac{1}{G}\right)\bigg/k_B T \qquad (12)$$

where $D$ is the diffusivity, $\Omega$ is the atomic volume, $k_B T$ has their usual meaning, and

$$G_{cr} = \frac{\sum G^3}{\sum G^2} \qquad (13)$$

It gives the steady-state growth kinetics as

$$\frac{d\bar{G}^3}{dt} = \left(\frac{4}{9}\right)(2D\Omega\gamma/k_B T) \qquad (14)$$

and the normalized steady-state size distribution $P_{eq}(x)$ as

$$P_{eq}(x) = P_{eq}\left(\frac{G}{\bar{G}}\right) = 3^4 \cdot 2^{-\frac{5}{3}} \cdot e \cdot x^2 (x+3)^{-\frac{7}{3}} \left(\frac{3}{2} - x\right)^{-\frac{11}{3}} \exp\left(-\frac{1}{1-2x/3}\right) \qquad (15)$$

Following the above approach, we obtain

$$\bar{G} = \frac{\int G P\left(\dfrac{G}{\bar{G}},t\right) dG}{\int P\left(\dfrac{G}{\bar{G}},t\right) dG} = \int G P\left(\dfrac{G}{\bar{G}},t\right) dG \qquad (16)$$

$$\frac{d\bar{G}^3}{dt} = 3\bar{G}^2 \frac{d\bar{G}}{dt}$$

$$= 3\bar{G}^2 \left[ -\left( GP\left(\frac{G}{\bar{G}},t\right)\frac{dG}{dt} \right)\bigg|_{G_{\min}}^{G_{\max}} + \int P\left(\frac{G}{\bar{G}},t\right)\frac{dG}{dt}dG + \int GP\left(\frac{G}{\bar{G}},t\right)dG \left( P\left(\frac{G}{\bar{G}},t\right)\frac{dG}{dt} \right)\bigg|_{G_{\min}}^{G_{\max}} \right]$$

(17)

By letting $P\left(\frac{G}{\bar{G}},t=0\right) = P_{eq}\left(\frac{G}{\bar{G}}\right)$, we obtain $\left.\frac{d\bar{G}^3}{dt}\right|_{t=0} = 0.444\left(2D\Omega\gamma/k_B T\right)$, which is consistent with the LSW solution of $\left(\frac{4}{9}\right)\left(2D\Omega\gamma/k_B T\right)$.

The last example is for 3-grain line controlled grain growth, where the coarsening equation can be written as

$$\frac{dG}{dt} = 2M_t \gamma \frac{G}{a}\left(\frac{1}{G_{cr}} - \frac{1}{G}\right) \qquad (18)$$

where $M_t$ is the mobility for 3-grain line, $a$ is the atomic size, and

$$G_{cr} = \frac{\sum G^4}{\sum G^3} \qquad (19)$$

Our analytical solution [5] shows the steady-state growth kinetics as

$$\frac{d\bar{G}}{dt} = \frac{1}{3}\left(\frac{2M_t \gamma}{a}\right) \qquad (20)$$

and the steady-state size distribution $P_{eq}(x)$ as

$$P_{eq}(x) = P_{eq}\left(\frac{G}{\bar{G}}\right) = \exp(-x) \qquad (21)$$

Following the same approach, we obtain

$$\bar{G} = \frac{\int GP\left(\frac{G}{\bar{G}},t\right)dG}{\int P\left(\frac{G}{\bar{G}},t\right)dG} = \int GP\left(\frac{G}{\bar{G}},t\right)dG \qquad (22)$$

$$\frac{d\bar{G}}{dt} = \left[ -\left( GP\left(\frac{G}{\bar{G}},t\right)\frac{dG}{dt}\right)\bigg|_{G_{\min}}^{G_{\max}} + \int P\left(\frac{G}{\bar{G}},t\right)\frac{dG}{dt}dG + \int GP\left(\frac{G}{\bar{G}},t\right)dG\left( P\left(\frac{G}{\bar{G}},t\right)\frac{dG}{dt}\right)\bigg|_{G_{\min}}^{G_{\max}} \right]$$

(23)

By letting $P\left(\frac{G}{\bar{G}},t=0\right) = P_{eq}\left(\frac{G}{\bar{G}}\right)$, we obtain $\frac{d\bar{G}}{dt}\bigg|_{t=0} = \frac{1}{3}\left(\frac{2M_t\gamma}{a}\right)$, which is the same as our analytical solution of $\frac{1}{3}\left(\frac{2M_t\gamma}{a}\right)$.